\documentstyle[12pt]{article}

\newcommand{\be}{\begin{equation}}
\newcommand{\ee}{\end{equation}}

\newcommand{\ba}{\begin{eqnarray}}
\newcommand{\ea}{\end{eqnarray}}

\newcommand{\D}{\Delta}

\newcommand{\e}{\epsilon}

\begin{document}

\hsize36truepc\vsize51truepc
\hoffset=-.4truein\voffset=-0.5truein
\setlength{\textheight}{8.5 in}

\begin{titlepage}
\begin{center}
\hfill LPTENS-00/25\\

\vskip 0.6 in
{\large  Interactions of several replicas in the random field Ising model}
\vskip .6 in
       {\bf Edouard Br\'ezin \footnote{$  ${\it
 Laboratoire de Physique Th\'eorique de l'\'Ecole Normale
Sup\'erieure, Unit\'e Mixte de Recherche 8549 du Centre National de la
Recherche
Scientifique et de l'\'Ecole Normale Sup\'erieure,
24 rue Lhomond, 75231 Paris Cedex 05, France.\\ {\bf
brezin@physique.ens.fr}}}}\ ,
            \hskip 0.3cm
       {\bf {Cirano De Dominicis}}  \footnote{ $  ${\it{ Service de
Physique Th\'eorique,CE Saclay\\ 91191 Gif-sur-Yvettte Cedex, France.
\\ {\bf cirano@spht.saclay.cea.fr}}}}\\
\end{center}

      \vskip 0.6in
{\bf Abstract}

\vskip 14.5pt

{\leftskip=0.5truein\rightskip=0.5truein\noindent
{\small

The replicated field theory of the random field Ising model involves
the couplings of replicas of different indices. The resulting correlation
functions involve
 a superposition of different types of  long distance behaviours. However
the $n=0$ limit allows one to
discuss the renormalization group properties in spite of this phenomenon.
The attraction of pairs of replicas
 is enhanced under renormalization flow and no stable
fixed point is found. Consequently an instability occurs in the
paramagnetic region, before
one reaches the Curie line, signalling the onset of replica symmetry breaking.

}
\par}

\newpage

\end{titlepage}
\setlength{\baselineskip}{1.5\baselineskip}


\section{  Introduction }

   In a recent article we have briefly examined the field theory associated
with
the random field
Ising model\cite{BDD, BDD2}, within the replica approach. It is a $\phi^4$
field theory, but with the following modifications :\\
 (i) there are $n$ fields $\phi_a$ , $a=1,\cdots,n$ and the only symmetry
is the permutation of the replicas. This
allows for five couplings namely $\displaystyle u_1\sum_a \phi_a^4 ,u_2
\sum_{ab} \phi_a^3 \phi_b, u_3\sum_{ab} \phi_a^2\phi_b^2,\\
u_4\sum_{abc} \phi_a^2\phi_b \phi_c,u_5 (\sum_1^n \phi_a)^4$.\\
(ii) in the $n=0$ limit the (bare) propagator at $T_c$ is given by
\be G_{ab}(p) = \frac{\delta_{ab}}{p^2} +\frac {\Delta}{(p^2)^2}.\ee
The second term results from the averaging over the random field $h(x)$ :
\be <h(x)h(y)> = \Delta \delta(x-y)\ee
At first sight, the $1/p^4$ singularity of the propagator could imply that
the upper critical dimension is eight,
instead of six, but it will be argued that it is indeed six because of the
$n=0$ limit. Then the renormalization group studies
which were conducted long ago \cite {{AIM},{Y},{PS}} dealt only with the
single coupling constant $u_1$, and one $\Delta/p^4$ propagator per loop
[contributions
with less than one $\Delta/p^4$ per loop being in any case infra-red
subdominant].
This limitation  was the result of considering the averaging over the
random field of connected correlation functions. However once
one introduces the coupling constants $u_2,\cdots,u_5$, which involve
several replicas, one must also consider contributions with moore tan one
$\Delta/p^4$ per loop. This modifies notably the
conventional renormalization programme, and we shall carry it in some detail
here in the zero replica limit. At the end we  recover the $n=0$ limit of the
beta-functions found in \cite{BDD} and thus confirm its main conclusion,
namely the instability of the dimensional reduction fixed point.

Furthermore we are left with infra-red singular and attractive
contributions to the 4-point
function of type $u_3$, i.e. corresponding to two distinct replicas. They
can give rise to negative eigenvalues in the iteration
of the related Bethe-Salpeter kernel, implying the occurence of a glassy
phase  in the paramagnetic domain, before one reaches
the Curie line \cite{MY}. We shall briefly return to this point at the end.

\section{  Upper critical dimension}
Let us first write the Boltzmann weight for this replicated field theory :
\ba\label{H} \beta H = \int d^d x \{ \frac{1}{2} \sum_a{[(\nabla \phi_a)^2
+ r_o(\phi_a)^2] -\frac{\Delta}{2}}\sum_{ab}\phi_a \phi_b  \nonumber\\
+ \frac{u_1}{4!} \sum_a \phi_a^4 + \frac{u_2}{3!} \sum_{ab} \phi_a^3 \phi_b
+ \frac{u_3}{8}\sum_{ab} \phi_a^2\phi_b^2 + \frac{
u_4}{4}\sum_{abc} \phi_a^2 \phi_b \phi_c + \frac{u_5}{4!} (\sum_a \phi_a)^4
\}. \ea
Indeed in \cite {BDD} it was argued that the standard derivation of a
field theory from the spin model in the presence of
a random field, namely the consideration of the fluctuations around mean
field theory, did yield those five coupling constants.
The corresponding propagator at the critical temperature  is thus simply
given by
\be G_{ab}(p) = \lim_{n \to 0} (\frac{\delta_{ab}}{p^2} +
\frac{\Delta}{p^2(p^2-n\Delta)}). \ee
In view of this $\Delta/p^4$ one does get infrared singularities in
dimensions lower than eight. For instance there is an obvious
one-loop contribution to the renormalization of $u_3$ proportional to
$(\Delta u_1)^2$ which is singular at  low external momenta as $1/p^{8-d}$.
However if one considers higher loops they are either less singular or they
vanish with $n$. Indeed the diagrams which would be maximally singular
in eight dimensions contain a $\Delta/p^4$ on every internal line.
Therefore as soon as the diagram contains any vertex which is not connected
to one or two of the four  external
lines, which is bound to happen at most at order five in perturbation
theory, it contains at least one free sum over replica indices, and thus it
vanishes with
$n$.

Therefore the singularities that one encounters in these functions between
eight and six dimensions are due to a finite number of graphs and
thus there is no possibility of anomalous dimensions,  which could modify
the behavior given by  simple dimensional analysis. This is very much like,
say a
six-point function at criticality, in a $\phi^4$ theory : it is singular at
low momentum for $d<6$ but this singularity remains canonical down to
four dimensions. Here similarly these singularities remain given by these
few graphs down to $d=6$, at which a full renormalization analysis becomes
 necessary : the upper critical dimension is six, not eight, because of the
$n=0$ limit.
\section{  Renormalization of the coupling constants }
Given the effective Hamiltonian (\ref{H}), we wish now to renormalize the
four-point functions, corresponding
to the five coupling constants $u_j (j=1,\cdots,5)$, retaining terms
involving one $\Delta$-propagator per loop
or more. We consider the one-particle-irreducible four point function
$\Gamma^{(4)}_{abcd} (p_1,\cdots,p_4)$ and in order to minimize the
number of momenta involved, we choose for simplicity the symmetric point
\be \label{SP}p_i\ . p_j = \frac{p^2}{3}(5\delta_{ij} -2) \ee
compatible with momentum conservation $\displaystyle \sum_1^4 p_i =0 $. We
work with  a dimensionally regularized
theory in dimension $d = 6-\epsilon$ , and will renormalize by minimal
subtraction.

For simplicity let us first keep only $u_1$ and $u_3$ alone and examine what
is happening.
The four-point function of type 1 , namely the one which involves a product
of three Kronecker deltas
 $\delta_{ab}\delta_{ac}\delta_{ad}$, is given , at one-loop order, by
\be \Gamma_1^{(4)}(p) = u_1 - 3u_1\frac{(\Delta
u_1)}{\epsilon}\frac{1}{p^{\epsilon}}+ \cdots . \ee
If we mutiply it by $\Delta$, one sees that the real coupling constants
which enter into all the diagrams which
involve exactly one $\Delta$ per loop are
\be g_i = \Delta u_i .\ee
Their (inverse length) dimension is $ \epsilon = 6-d$.
Therefore in a minimal scheme, the relation between the bare $g_1$ and the
dimensionless
renormalized $g_1^R$ is, at this order
\be g_1 = \mu^{\e} g_1^R [ 1 + 3\frac{g_1^R}{\e}  +O((g_1^R)^2)]. \ee
Therefore the  corresponding contribution to the beta function is
\be \beta_1 = -\e g_1^R + 3 (g_1^R)^2 + \cdots.\ee
For the four-point function of type 3 , namely the one which involves a
product of two Kronecker deltas of the form
 $[\Gamma^{(4)}_3]_{ab}\left(\delta_{ad}\delta_{bc}+ \delta_{ac}\delta_{bd}
+\delta_{ab}\delta_{cd}\right)$, the situation is
different since we now meet two kinds of diagrams, those with one $\D$
perloop and those in which
the number of $\D$'s is equal to the number of loops plus one. On purely
dimensional grounds we may decompose
$\D \Gamma_3^{(4)}$ as
\be \D \Gamma_3^{(4)}(p) = \gamma_3(p) + \delta_3(p) \ee
with
\ba \gamma_3(p) &=& g_3 -  2 \frac{g_1g_3}{\e}\frac{1}{p^{\e}} +\cdots
\nonumber\\
 \delta_3(p) &=& \frac{\D}{p^{2+\e}}g_1^2 +\cdots \ea
(i) Consider first the most infra-red singular contribution $\delta_3(p)$.
It comes from graphs
with one $\Delta$ per loop,plus one. And its vertices can only be
$g_1$'s, since the occurence of one $g_3$ or more ( like the insertion of more
$\Delta$'s) would entail some free replica summations vanishing with $n$.
Furthermore
the structure of those diagrams in $[\delta_3(p_1,p_2,p_3, p_4)]_{ab}$ is
of the form
\ba [\Gamma^{(4)}_3]_{ab}(p_1,\cdots,p_4) = (1-\delta_{ab})\D^2\int d^dq \
\Gamma^{(4)}_1(p_1,p_2,q, p_1+p_2-q)
\nonumber\\\times G_{ab}(q)G_{ab}(p_1+p_2-q)
\Gamma^{(4)}_1(q, p_1+p_2-q,p_3,p_4)+ \rm{permutations},\ea
in which $G_{ab}(q)$ is the renormalized propagator. The integral over $q$
is convergent,
since $(1-\delta_{ab})$ selects the part fo the propagator which falls off as
$1/q^4$ (up to logarithms).
Therefore the renormalization of $\Gamma^{(4)}_1$ and of the propagators
is sufficient to make
the diagrams contributing to $\delta_3(p)$ finite. No new renormalization
is needed, and $\delta_3(p_1,p_2,p_3, p_4)$ satisfies a Callan-Symanzik
equation
per se. \\
(ii) $\gamma_3(p)$ comes from graphs with excatly one $\Delta$ per loop.
They are
all linear in $g_3$ since any higher power would again lead to free
replica summations vanishing with $n$.
 Therefore the minimal
renormalization of $g_3$, defined as
\be g_3 =  \mu^{\e} g_3^R [ 1 + 2\frac{g_1^R}{\e}  +O((g_1^R)^2)]. \ee
is sufficient to make $\gamma_3(p)$ finite. The beta-function for $g_3$
follows :
\be \beta_3 = -\e g^R_3 + 2 g^R_1g^R_3 + O(g^R_3 (g^R_1)^2) .\ee

In general for the theory with the five coupling constants the same pattern
governs the renormalization procedure. The functions $\Gamma^{(4)}_2$ and
$\Gamma^{(4)}_3$ will both involve also terms with one
$\D$ per loop, linear in $g_2$ and $g_3$,  which  lead to a renormalization
of $g_2$ and $g_3$ ; they have both contributions
with one more $\D$ which are made finite by the previous renormalization of
$\Gamma^{(4)}_1$. For $\Gamma^{(4)}_4$, one finds first terms linear in
$g_4$ and quadratic in $g_2$ and $g_3$ which
come from one $\D$ per loop ; they lead to a renormalization of $g_4$. Then
one finds terms proportional to $\D/p^2$,
linear in $g_2$ and $g_3$,  and terms proportional to $(\D/p^2)^2$, which
are made finite by the previous renormalization of
$\Gamma^{(4)}_1$ and of the propagator.  For $\Gamma^{(4)}_5$
 the situation is again similar except that there are now terms up to
$(\D/p^2)^3$ namely with $k$ more  $\D$'s than the number of loops,
$k= 0,\cdots,3$. The $k=0$ terms lead to a renormalization of $g_5$ ; the
other ones are finite as a consequence of previous
renormalizations.
The five beta functions at one-loop order are then \cite{BDD} :
\ba \beta_1 &=& -\e g_1 + 3 g_1^2\nonumber \\
\beta_2 &=& -\e g_2 + 3 g_1 (g_2+g_3)\nonumber \\
\beta_3 &=& -\e g_3 + 2 g_1 (g_2+g_3)\nonumber \\
\beta_4 &=& -\e g_4 + 3 g_1 g_4 + 4(g_2+g_3)^2\nonumber \\
\beta_5 &=& -\e g_5 + 36 g_4 (g_2+g_3). \ea
The dimensional reduction fixed point, namely $g_1 = \frac{1}{3} \e+
O(\e^2)$ and $g_2= \cdots= g_5= 0$ is unstable.
(It is sufficient to notice that at this fixed point the matrix of
derivatives $\frac{\partial \beta_i}{\partial g_j}$ has an eigenvalue
equal to $ \frac{\partial \beta_3}{\partial g_3} = -\frac{1}{3}\e $).

\section{  Field renormalization }
At two-loop order a wave function renormalization appears. However a priori
the random field  introduces a privileged
direction in the internal space along the unit vector $\vec{v} =
\frac{1}{\sqrt n}(1,\cdots,1)$. We have thus to introduce the longitudinal
and tranverse components of the fields
\be \phi_L = \vec \phi . \vec v = \frac{1}{\sqrt n}\sum_{a=1}^n \phi_a \ee
and
\be \vec \phi_T = \vec \phi - \phi_L \vec v. \ee
A priori they are both renormalized by different factors and we define the
renormalized fields through
\be \phi_L = \sqrt{Z_L} \Phi_L , \   \vec \phi_T = \sqrt{Z_T} \vec\Phi_T , \ee
in which the $n$ fields $\Phi$ have finite correlation functions when
$\epsilon =( 6-d)$ goes to zero.
In terms of these fields the quadratic terms in the action have the form
\be \beta H_0 = \int d^d x \{ \frac{1}{2} {[(\nabla \phi_L)^2 + (\nabla
\vec\phi_T)^2] -n\frac{\Delta}{2}}\phi_L^2
\},\ee
i.e. in terms of the renormalized fields
\be \beta H_0 = \int d^d x \{ \frac{1}{2} {[Z_L(\nabla \Phi_L)^2 +Z_T
(\nabla \vec\Phi_T)^2] -nZ_L\frac{\Delta}{2}}\Phi_L^2
\}.\ee
This gives for the renormalized two-point function
\be \Gamma^{(2)}_{ab} (p) = Z_L (p^2-n\Delta) \frac{1}{n} + Z_T
p^2(\delta_{ab} -\frac{1}{n}) - \Sigma_{ab}, \ee
in which we have used the projectors along the vector $\vec v$, the
constant matrix whose elements are $1/n$, and the projector transverse to
$\vec v$,
the matrix
$\delta_{ab} - 1/n$ ; $\Sigma_{ab}$ contains all the self-energy diagrams.

From this expression one sees that it is expected that the wave function
renormalization
of the longitudinal part of the propagator $Z_L$  renormalizes as well
$\Delta$, the
variance of the random field. This is not obvious a priori; in the appendix
an explicit calculation
to two-loop order is given, which shows that to this order
$Z_T= Z_L$ and that $\Delta$ does not acquire an independent renormalization.
In a Langevin  dynamics of the same problem
\cite{BDD2} the non-renormalization of $\Delta$ is a natural consequence
of the non-renormalization of the temperature.
\section {Instability}
  A derivation of the Landau theory with the five $\phi^4$
coupling constants, was given in \cite{BDD}. It was shown there that the
coupling
constant $g_3= u_3\Delta$,
which couples two distinct replicas, was attractive. We
note that under renormalization
$g_3$ is pushed further away from the origin. It is thus natural to examine
whether
this attractive interaction may lead to an instability.  This question has
been considered in some details
by several authors \cite{MY,DA, MM, DD}. Here we wish to keep to very simple
arguments.

Therefore we now
consider whether the
Bethe-Salpeter kernel, for a pair of replicas of different indices, might
develop a
vanishing eigenvalue, thereby signaling bound state formation. Note that
couplings other
than $g_3$ vanish with $n$ under iteration in the channel of two (distinct)
replicas.
The simplest iterative kernel (besides $g_3$)  is a "bubble", i.e. a
$\delta_3$-like contribution followed by two
propagators, i.e.
\be \tilde{g}_3^2 \frac{1}{[(p-q)^2]^2} \int d^d k
\frac{1}{(k^2)^2[(k-q)^2]^2}. \ee
An instability takes place whenever the spectrum of the operator
\be\label{Schr} \hat{h} \psi = p^4 \psi(p)- \tilde{g}_3^2\int d^d q  B(p-q)
\psi(q) \ee
has a vanishing eigenvalue, in which
\be B(p) = \int d^d k \frac{1}{(k^2)^2[(k-p)^2]^2}.\ee
In position space this operator is
\be \hat{h} = [(\nabla)^2]^2 - C\tilde{g}_3^2  \frac{1}{r^{2d-8}} \ee
in which $C$ is a positive constant.
The four derivatives may balance the singularity at the origin of the
attractive potential whenever $2d-8\leq 4$. In that range the spectrum
consists
of bound states and positive energy scattering states. A simple scaling
shows that the binding energies, in the domain $d\leq6$ are proportional to
$\displaystyle\mid\tilde{g}_3\mid ^{\frac{4}{6-d}}$. Therefore whenever
$\displaystyle\mid\tilde{g}_3\mid ^{\frac{4}{6-d}}$ is larger than some
critical value a
bound state at zero energy appears, signaling an instability.
Since $\tilde{g}_3$ is proportional to $\Delta $, it means that, for
$\Delta$ small,
there is a small domain of size proportional to $\Delta^{4/(6-d)}$ above
the Curie line, in which the attraction between pairs of distinct replicas
generate an instability.

 Quite generally it can be shown that the $\hat{h}$ operator of
(\ref{Schr}) is, beyond
the one-loop approximation, the Jacobian matrix of the Legendre transform
\cite{DD}
with respect to a source $\Delta_{ab}(p)$
\be W(\Delta_{ab}) + \Gamma (G_{ab}) = \frac{1}{2}\sum_p
\sum_{ab}\Delta_{ab}(p)
 G_{ab}(p)\ee
i.e.
\ba \hat{h}_{ab}(p;p')& =& \frac{\partial^2 \Gamma}{\partial
G_{ab}(p)\partial G_{ab}(p')}\nonumber\\
&=& [G_{aa}(p)]^{-1}[G_{bb}(p)]^{-1} \delta_{p+p';0} -
\Gamma^{4}_3(p,-p;p',-p').\ea
On the other hand that double derivative has been identified with the
replicon component
of the Hessian matrix \cite{MY,DD}, around the replica symmetric
solution for the connected ($
G_{aa} =G$) and disconnected ( or connected through the random averaging,
$G_{ab} = \tilde{G}$)
components. Thus the instability, i.e. negative eigenvalues of $\hat{h}$,
will force the emergence of a replica symmetry
broken solution \cite{MY,DA}.

To conclude in one sentence, it may be said that the breakdown of
dimensional reduction is not simply that
the $\epsilon$-expansion needs to be corrected by non-analytic terms, but
follows from a change in the phase diagram itself with the emeregence of a
new, glassy, phase.

\vskip 8mm
\setcounter{equation}{0}
\renewcommand{\theequation}{A.\arabic
{equation}}
{\bf {Appendix} : {The propagator at two-loop order}}
\vskip 5mm
The contributions to the self-energy due to the coupling constant $g_1$ vanish
at one-loop in the minimal subtraction scheme. At two-loop the singular
terms of the self-energy are given by
\be \Sigma_{ab} = \tilde{g}_1^2 [ A(p) \delta_{ab} + \Delta B(p) ]\ee
with
\be A(p) = \frac{1}{2}\int d^dq_1d^dq_2
\frac{1}{q_1^4}\frac{1}{q_2^4}\frac{1}{(p+q_1+q_2)^2}\ee
and
\be B(p) = \frac{1}{6}\int d^dq_1d^dq_2
\frac{1}{q_1^4}\frac{1}{q_2^4}\frac{1}{(p+q_1+q_2)^4},\ee
\be \tilde{g}_1 = g_1 \frac{2\pi^{d/2}}{(2\pi)^d\Gamma(d/2)}.\ee
A lengthy, but standard, dimensional calculation in which $d=6-\epsilon$,
leads to
\ba A &=& -\frac {p^{2+2\e}}{12 \e} \frac{(1-\e/2)^2 (1-\e/4)^2}{(1-3\e/2)
(1-3\e/4)}\frac{\Gamma^5(1-\e/2)\Gamma(1+\e)}
{\Gamma(1-3\e/2)}\nonumber\\ &=& - \frac{p^2}{12\e} + \rm{finite}\ea
 and
\ba B &=& \frac {p^{-2\e}}{12 \e} \frac{(1-\e/2)^2 (1-\e/4)^2
(1+\e)}{(1+\e/2) (1+\e/4)(1-\e)}\frac{\Gamma^5(1-\e/2)\Gamma(1+\e)}
{\Gamma(1+\e/2)\Gamma(1-\e)}\nonumber\\ &=& \frac{1}{12\e} + \rm{finite}.\ea
At this order the one-particle irreducible two-point function is  given by
\be \Gamma^{(2)}_{ab}(p) = Z_T(\delta_{ab} - \frac{1}{n}) p^2 + Z_L
\frac{1}{n} (p^2 - n \Delta)  - \Sigma_{ab} \ee
and thus in the minimal subtraction scheme, we find that, since $A$ and $B$
diverge with the same $\pm\frac{1}{12\e}$
\be Z_T = Z_L = 1 + \frac{ \tilde{g}_1^2}{12 \e} + O(\tilde{g}_1^3),\ee
and that indeed $\Delta$ does not require any additional renormalization.

\end{document}